\documentclass[letterpaper, 10 pt, conference]{ieeeconf}  

\IEEEoverridecommandlockouts                              
\usepackage{amssymb}
\usepackage{stfloats}
\usepackage[dvips]{color}
\usepackage[dvips]{graphicx}
\usepackage{amsmath}
\usepackage{multirow,subfigure}
\usepackage{bm}
\usepackage{booktabs}

\title{\LARGE \bf
Approximate Interval Method for Epistemic Uncertainty Propagation using Polynomial Chaos and Evidence Theory
}

\author{Gabriel Terejanu, Puneet Singla, Tarunraj Singh, Peter D. Scott
\thanks{This work was supported under Contract No. HM1582-08-1-0012 from ONR.}
\thanks{Gabriel Terejanu - Ph.D. Candidate, Department of Computer Science \& Engineering, University at Buffalo, Buffalo, NY-14260, {\tt\small terejanu@buffalo.edu}}%
\thanks{Puneet Singla - Assistant Professor, Department of Mechanical \& Aerospace Engineering, University at Buffalo, Buffalo, NY-14260, {\tt\small psingla@buffalo.edu}}%
\thanks{Tarunraj Singh - Professor, Department of Mechanical \& Aerospace Engineering, University at Buffalo, Buffalo, NY-14260, {\tt\small tsingh@buffalo.edu}}%
\thanks{Peter D. Scott - Associate Professor, Department of Computer Science \& Engineering, University at Buffalo, Buffalo, NY-14260, {\tt\small peter@buffalo.edu}}%
}

\begin{document}

\maketitle
\thispagestyle{empty}
\pagestyle{empty}

\begin{abstract}
The paper builds upon a recent approach to find the approximate bounds of a real function using Polynomial Chaos expansions. Given a function of random variables with compact support probability distributions, the intuition is to quantify the uncertainty in the response using Polynomial Chaos expansion and discard all the information provided about the randomness of the output and extract only the bounds of its compact support. To solve for the bounding range of polynomials, we transform the Polynomial Chaos expansion in the Bernstein form, and use the range enclosure property of Bernstein polynomials to find the minimum and maximum value of the response. This procedure is used to propagate Dempster-Shafer structures on closed intervals through nonlinear functions and it is applied on an algebraic challenge problem.
\end{abstract}

\section{INTRODUCTION}
\label{sec:Introduction}
In interval analysis a fundamental problem is finding the interval bounds for the range of a real function. When such a function is monotone or it can be expressed in terms of arithmetic operations, then interval computations can be used to approximate the bounds of the response. However these bounds are gross overestimations due to the dependency and the wrapping effect \cite{Jaulin2001}. 

Applications of interval methods can be found in estimation, optimization techniques, robust control, robotics and finance \cite{Moore1979, Jaulin2001}, just to name a few. First, introduced by Moore \cite{Moore1966}, as a method to control for numerical errors in computers, the field of interval analysis has evolved with better approximations to the range of real functions as presented in Ref.\cite{Ratschek1984, Rall1985}.

In the present paper we are interested in using interval methods to propagate \textit{epistemic uncertainty} through nonlinear functions. In contrast to the \textit{aleatory uncertainty} \cite{Terejanu2008, Terejanu2009a, Konda2009}, defined by variability which is irreducible, the epistemic uncertainty is derived from incomplete knowledge or ignorance and can be reduced with an increase in information. Due to their major differences, it is of great importance that the two type of uncertainties to be modeled and propagated separately \cite{Ferson1996}.

Unlike the probability theory, where the probability mass is assigned to singletons, in the Shafer's theory of evidence \cite{Shafer1976}, the probability mass is assigned to sets, given its power on modeling ignorance. In conjunction with the Dempster's rule of combinations \cite{Dempster1967} which is a generalization of the of the Bayes' rule, the Dempster-Shafer (DS) theory of evidence offers a powerful methodology for representing and aggregating epistemic uncertainties. 

One can define Dempster-Shafer structures on focal elements that are closed intervals on the real line for example. Ferson \cite{Ferson2003} shows how this structures can be transformed into probability bounds and vice-versa by discretization. To propagate these focal elements through system functions, it involves finding the solution to the interval propagation problem.

We build upon a recent approach to find the approximate bounds of a real function using Polynomial Chaos expansions introduced by Monti \cite{Monti2008,Smith2007} and applied for worst-case analysis and robust stability. Given a function of random variables with compact support probability distributions, the intuition is to quantify the uncertainty in the response using Polynomial Chaos expansion and discard all the information provided about the randomness of the output and extract only the bounds of its compact support. 

Introduced by Norbert Weiner \cite{Weiner1938}, Polynomial Chaos initially coined as the Homogeneous Chaos was used to represent a Gaussian process as a series of Hermite polynomials. This method has been generalized to the Askey-scheme of orthogonal polynomials used to model random variables characterized by different probability density functions, including Beta and Uniform which have compact support \cite{Xiu2002}. 

The Polynomial Chaos is mathematically attractive due to the functional representations of the stochastic variables. It separates the deterministic part in the polynomial coefficients and the stochastic part in the orthogonal polynomial basis. This becomes particularly useful in characterizing the uncertainty of the response of a dynamical system represented by ordinary differential equations with uncertain parameters.

To solve for the bounding range of polynomials, we propose to transform the Polynomial Chaos expansion in the Bernstein form, and use the range enclosure property of Bernstein polynomials to find the minimum and maximum value of the response \cite{Cargo1966}. The transformation does not require polynomial evaluations and it guarantees the global optimality of the bounds \cite{Garloff1985} and it is shown to be more efficient than existent interval global optimizers \cite{Ray2008}.

To demonstrate this approach for propagating Dempster-Shafer structures on closed intervals through nonlinear functions, we apply the proposed method on an algebraic challenge problem used to investigate the propagation of  epistemic uncertainty \cite{Oberkampf2004}.

The problem of propagating DS structures on closed intervals through nonlinear functions is stated in Section~\ref{sec:Problem} and the proposed method is presented in Section \ref{sec:Method}. The numerical  example is given in Section~\ref{sec:Results} and the conclusions and future work are discussed in Section~\ref{sec:Conclusions}.

\section{PROBLEM STATEMENT}
\label{sec:Problem}
\subsection{Theory of Evidence}
\label{sec:Evidence}
The primitive function in the theory of evidence is the \textit{basic probability assignment (bpa)}, represented here by $m$, which is similar to the probability in the probability theory.  The bpa for a given set can be understood as the weight of evidence that the truth is in that set, evidence, which cannot be further subdivided among the members of the set. The bpa defines a map of the power set over the \textit{frame of discernment} $\Omega$ to the interval $[0,1]$: $m:2^\Omega \rightarrow [0,1]$. The \textit{focal element} of $m$ is every subset $A \subseteq \Omega$ such that $m(A)>0$ and the \textit{belief structure} $m$ verifies:
\begin{eqnarray}
\sum_{A \subseteq \Omega} m(A) = 1 \quad \mathrm{where} \quad m(A_i) = p_i
\end{eqnarray} 

In this work we are considering normalized belief structures (closed-world assumption) which satisfy the following relation: $m(\phi)=0$, where $\phi$ is the null set. As an example consider the following \textit{body of evidence} $(\Omega, m)$: $\Omega=\{M,N,P\}$ with $m(\{M,N\})=0.3$ and $m(\{N,P\})=0.7$.

Based on the mass function two new functions can be induced. The \textit{belief function} or the \textit{lower bound}, $Bel$, which quantifies the total amount of support given to the set of interest $A$:
\begin{eqnarray}
Bel(A) = \sum_{B \subseteq A} m(B)
\end{eqnarray} 	

The \textit{plausibility function} or the \textit{upper bound}, $Pl$, which quantifies the maximum amount of potential given to the set of interest $A$ (here $\bar{A}$ is the complement of $A$):
\begin{eqnarray}
Pl(A) = \sum_{B \cap A \neq \phi} m(B) = 1 - Bel(\bar{A})
\end{eqnarray} 	

The precise probability is bounded by the two quantities defined above, and when the equality is satisfied then the belief measure is just a probability measure and all the focal elements are singletons.
\begin{eqnarray}
Bel(A) \le prob(A) \le Pl(A)
\end{eqnarray} 	

Given two bpa's $m_1$ and $m_2$ based on independent arguments on the same frame of discernment, the Dempster's rule of combination provides the means to calculate the aggregation of the two belief structures:
\begin{eqnarray}
m_{12}(A \neq \phi) &=& \frac{1}{1-K} \sum_{B \cap C = A} m_1(B) m_2(C) \\
m_{12}(\phi) &=& 0 \nonumber
\end{eqnarray} 	
where $K = \sum_{B \cap C = \phi} m_1(B) m_2(C)$ represents the amount of probability mass due to conflict.

While a number of combination rules have been derived to aggregate information \cite{Sentz2002}, we present also the mixing rule which is used in the numerical example. Given $n$ belief structures to be aggregated, the formula for the mixing rule is given by:
\begin{eqnarray}
m_{1\ldots n}(A) &=& \frac{1}{n}\sum_{i=1}^n w_i m_i(A) \label{mixingRule}
\end{eqnarray}
where $w_i$ are the corresponding weights proportional with the reliability of the sources.
\subsection{DS structures on closed intervals}
\label{sec:DSstructures}
Given two nondecreasing functions $\overline{F}$ and $\underline{F}$, where $\overline{F}, \underline{F} : \mathbb{R} \rightarrow [0,1]$ and $\underline{F}(x) \le \overline{F}(x)$ for all $x \in \mathbb{R}$, we can represent the imprecision in the cumulative distribution function (CDF), $F(x)=Prob(X\le x)$, by the probability box (p-box) $[\underline{F},\overline{F}]$ as follows: $\underline{F}(x) \le F(x) \le \overline{F}(x)$ \cite{Ferson2003}.

A Dempster-Shafer structure on closed intervals can induce a unique p-box, while the inverse is not uniquely determined. Many Dempster-Shafer structures exist for the same p-box. Given the following body of evidence, $\big\{~([\underline{x}_1,\overline{x}_1],p_1) ~,~ ([\underline{x}_2,\overline{x}_2],p_2) ~,~ \ldots ~([\underline{x}_n,\overline{x}_n],p_n) ~\big\}$, the cumulative belief function (CBF) and the cumulative plausibility function (CPF) are defined by:
\begin{eqnarray}\label{cdf}
CBF(x) = \underline{F}(x) = \sum_{\overline{x}_i \le x} p_i \\
CPF(x) = \overline{F}(x) = \sum_{\underline{x}_i \le x} p_i 
\end{eqnarray} 

Similarly one can obtain the complementary cumulative belief function (CCBF) and the complementary cumulative plausibility function (CCPF).
\begin{eqnarray}
CCBF(x) = 1 - CPF(x) =  \sum_{\underline{x}_i > x} p_i \label{ccdf1}\\
CCPF(x) = 1 - CBF(x) = \sum_{\overline{x}_i > x} p_i \label{ccdf2}
\end{eqnarray} 

Thus the complementary cumulative distribution function (CCDF), $F_c(x) = Prob(X > x)$, is bounded as follows:
\begin{eqnarray}
CCBF(x) \le F_c(x) \le CCPF(x)
\end{eqnarray} 

\begin{figure}[h]
	\centering
		\includegraphics[width=2.5in]{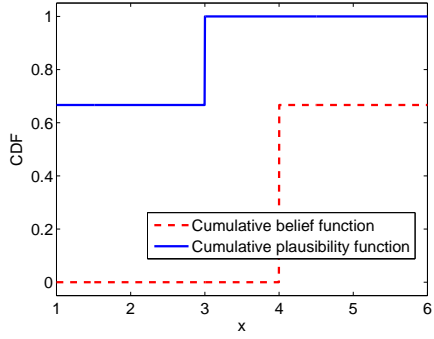}
	\caption{P-box induced by the Dempster-Shafer structure}
	\label{fig:CDF_interval_simple}
\end{figure}

\textit{Example}: Consider the following body of evidence $\big\{~([1,4],2/3) ~,~ ([3,6],1/3) ~\big\}$, the lower and the upper cumulative functions are plotted in Fig.\ref{fig:CDF_interval_simple}.
\subsection{Mapping of DS structures}
\label{sec:Mapping}
Consider the following function: $y = f(a,b)$, where $f : \mathbb{R}^2 \rightarrow \mathbb{R}$, and $a$ and $b$ are given by the following bodies of evidence $\big\{~([\underline{a_1},\overline{a_1}],p_1^a)  ~,~ \ldots ~([\underline{a_n},\overline{a_n}],p_n^a) ~\big\}$ and $\big\{~([\underline{b_1},\overline{b_1}],p_1^b)  ~,~ \ldots ~([\underline{b_n},\overline{b_n}],p_n^b) ~\big\}$ respectively. We are interested in finding the induced Dempster-Shafer structure in the $y$ variable. The basic probability assignment describing $y$ is given by \cite{Yager1986}:
\begin{equation}\label{mapping}
m_f(Y) = \sum_{f(A_i, B_j) = Y} \underbrace{m_1(A_i)}_{p_i^a} \underbrace{m_2(B_j)}_{p_j^b}
\end{equation}
where $Y=[\underline{y},\overline{y}], A_i = [\underline{a_i},\overline{a_i}]$ and $B_j = [\underline{b_j},\overline{b_j}]$

Thus the problem of finding the mapping of a body of evidence on closed intervals is reduced to interval propagation \cite{Limbourg2008}. This problem can be solved using the advanced techniques developed in the interval analysis field \cite{Jaulin2001}. However, due to the dependency problem the obtained bounds are conservative which is detrimental to the belief structure, since the evidence is assigned automatically to other elements which are not in the body of evidence. This problem becomes more acute when the uncertainty has to be propagated over a period of time.

\section{PROPOSED APPROACH}
\label{sec:Method}
We propose a new approach in approximating the propagation of intervals using a non-intrusive polynomial chaos method \cite{Debusschere2005} in combination with Bernstein polynomials \cite{Stahl1995}. The approach of using polynomial chaos in propagating epistemic uncertainty has been considered previously in Ref.\cite{Monti2008} for a limited number of arithmetic operations and relies on sampling or global optimization in the general case which is inaccurate for small number of samples and computationally expensive.

\subsection{Non-intrusive polynomial chaos}
\label{sec:PolyChaos}
The problem in solving the mapping of DS structures as given in Eq.\eqref{mapping} is to find $Y=[\underline{y},\overline{y}]$ such that:
\begin{eqnarray}
Y = f(A, B) \label{IntProp}
\end{eqnarray}
where $ A = [\underline{a},\overline{a}]$ and $B = [\underline{b},\overline{b}]$. While in this paper we present only the bivariate case, the method can be scaled up to the desired number of variables. 

The problem can be transformed into finding the stochastic response $y$ by defining $a \sim \mathcal{U}(\underline{a},\overline{a})$ and $b \sim \mathcal{U}(\underline{b},\overline{b})$:
\begin{eqnarray}
y = f(a, b)
\end{eqnarray}
and write the polynomial chaos expansion for the uncertain arguments and the response:
\begin{eqnarray}
a = \sum_{i=0}^{p-1} a_i \psi_i(\xi_1) \quad \mathrm{where} \quad \xi_1 \sim \mathcal{U}(-1,1) \\
b = \sum_{j=0}^{p-1} b_j \psi_j(\xi_2) \quad \mathrm{where} \quad \xi_2 \sim \mathcal{U}(-1,1) \\
y = \sum_{k=0}^{P-1} y_k \bm{\psi}_k(\bm{\xi}) \quad \mathrm{where} \quad P = \frac{(n+p)!}{n!p!} \label{responsePC}
\end{eqnarray}

For this paper we are only concerned with the Uniform distribution, however due to the specificity of this application any other probability distribution with compact support can be used (eq. Beta). The sensitivity of the method with respect to the shape of the probability density function over the compact support remains to be studied.

Here $n$ is the number of uncertain input variables and $p$ the order of the polynomial chaos expansion. The basis function $\psi_n$ is the $n$-th degree Legendre polynomial and the polynomial coefficients of the input variables are given by:
\begin{eqnarray}
a_0 = \frac{\overline{a}+\underline{a}}{2} ~,~ a_1 = \frac{\overline{a}-\underline{a}}{2} ~,~ a_2 = \ldots = a_{p-1} = 0 \\
b_0 = \frac{\overline{b}+\underline{b}}{2} ~,~ b_1 = \frac{\overline{b}-\underline{b}}{2} ~,~ b_2 = \ldots = b_{p-1} = 0 
\end{eqnarray}

The first six multidimensional Legendre polynomials for the bivariate case are given by:
\begin{eqnarray}
\bm{\psi}_0(\bm{\xi}) &= \psi_0(\xi_1)\psi_0(\xi_2) &= 1 \\
\bm{\psi}_1(\bm{\xi}) &= \psi_1(\xi_1)\psi_0(\xi_2) &= \xi_1 \\
\bm{\psi}_2(\bm{\xi}) &= \psi_2(\xi_1)\psi_0(\xi_2) &= \frac{1}{2}(3\xi_1^2-1) \\
\bm{\psi}_3(\bm{\xi}) &= \psi_0(\xi_1)\psi_1(\xi_2) &= \xi_2 \\
\bm{\psi}_4(\bm{\xi}) &= \psi_1(\xi_1)\psi_1(\xi_2) &= \xi_1 \xi_2 \\
\bm{\psi}_5(\bm{\xi}) &= \psi_2(\xi_1)\psi_1(\xi_2) &= \frac{1}{2}(3\xi_1^2-1) \xi_2
\end{eqnarray}

We are interested in finding the polynomial coefficients $y_k$ which characterize the stochastic behavior of the output variable. Using the Galerkin projection and the orthogonality property of the polynomials one can isolate the coefficients $y_k$ as shown in Ref. \cite{Eldred2008}:
\begin{eqnarray} \label{Ycoeff}
y_k = \frac{<f,\bm\psi_k>}{<\bm\psi_k^2>} = \frac{1}{<\bm\psi_k^2>} \int_\Omega f \psi_k \varsigma(\bm\xi) \mathrm{d}\bm\xi
\end{eqnarray}
where $\varsigma(\bm\xi) = \prod_{i=1}^n \varsigma_i(\xi_i)$ is the joint probability density function. The integral can be evaluated using sampling or quadrature techniques.

We show that by bringing the polynomial chaos expansion to a Bernstein form using the Garloff's method \cite{Garloff1985}, we can efficiently find the minimum and the maximum value of the compact support thanks to the properties of the Bernstein polynomials: the smallest and the largest coefficient bound the output of the function modeled.

To transform our expansion from Legendre polynomial basis to Bernstein polynomial basis we expand our Polynomial Chaos expansion, Eq.\eqref{responsePC} into a simple power series and identify the new coefficients:
\begin{eqnarray}
y = \sum_{\mathbf{I} \le \mathbf{N}} \alpha_\mathbf{I} \bm\xi^\mathbf{I} \label{powerSeries}
\end{eqnarray}
where the multi-index $\mathbf{I} = (i_1, \ldots, i_n) \in \mathbb{N}^n$ and $\mathbf{N} = (n_1, \ldots, n_n) \in \mathbb{N}^n$ is the multi-index of maximum degrees; thus the maximum degree of $\xi_k$ is given by $n_k$. Here, we denote $\bm\xi^I = \xi_1^{i_1} \cdot \ldots \cdot  \xi_n^{i_n}$ and the inequality $\mathbf{I} \le \mathbf{J}$ implies $i_1 \le j_1, \ldots, i_n \le j_n$.

\subsection{Garloff's method to calculate the Bernstein coefficients}
\label{sec:Garloff}
We are interested in the transformation of the power series in Eq.\eqref{powerSeries} into its Bernstein form:
\begin{eqnarray}
y = \sum_{\mathbf{I} \le \mathbf{N}} \beta_\mathbf{I} \mathbf{B}_{\mathbf{I}}^{\mathbf{N}}(\bm\xi) \label{bernExpansion}
\end{eqnarray}
where $\mathbf{B}_{\mathbf{I}}^{\mathbf{N}}(\bm\xi)$ is the $\mathbf{I}$th Bernstein polynomial of degree $\mathbf{N}$ on the general box $G = [\underline{\bm\xi}, \overline{\bm\xi}]$. In our bi-variate case $G = [-1,1] \times [-1,1]$ since $\xi_1, \xi_2 \sim \mathcal{U}(-1,1)$.
\begin{eqnarray}
\mathbf{B}_{\mathbf{I}}^{\mathbf{N}}(\bm\xi) = B_{i_1}^{n_1}(\xi_1) \cdot \ldots \cdot B_{i_n}^{n_n}(\xi_n)
\end{eqnarray}

The univariate Bernstein polynomial $B_k^n(\xi)$ on the general interval $[\underline{\xi}, \overline{\xi}]$ is given by:
\begin{eqnarray}
B_k^n(\xi) = {n \choose k} \frac{(\xi-\underline{\xi})^k(\overline{\xi}-\xi)^{n-k}}{(\overline{\xi}-\underline{\xi})^n}
\end{eqnarray}

The Bernstein coefficients $\beta_{\mathbf{I}}$ are given by:
\begin{eqnarray}
\beta_{\mathbf{I}} = \sum_{\mathbf{J} \le \mathbf{I} \le \mathbf{N}} \frac{{\mathbf{I} \choose \mathbf{J}}}{{\mathbf{N} \choose \mathbf{J}}} \hat{\alpha}_{\mathbf{J}} \label{bernCoeff}
\end{eqnarray}
where we write ${\mathbf{I} \choose \mathbf{J}} = {i_1 \choose j_1} \cdot \ldots \cdot {i_n \choose j_n}$.

The scaled coefficients $\hat{\alpha}_{\mathbf{I}}$ are obtained as described in Ref.\cite{Berchtold2000} from the $\alpha_{\mathbf{I} }$ coefficients in Eq.\eqref{powerSeries} and the box $G$:
\begin{eqnarray}
\hat{\alpha}_{\mathbf{I}} &=& \tilde{\alpha}_\mathbf{I}(\overline{\bm\xi} - \underline{\bm\xi})^\mathbf{I} \label{finalCoeff} \\
\tilde{\alpha}_\mathbf{I} &=& \sum_{\mathbf{I} \le \mathbf{J} \le \mathbf{N}} {\mathbf{J} \choose \mathbf{I}} \alpha_\mathbf{J} \underline{\bm\xi}^{\mathbf{J}-\mathbf{I}} \label{interCoeff}
\end{eqnarray}

\textit{Example}: Consider the following Polynomial Chaos expansion given by $n=2, p=3$ and $P=10$:
\begin{eqnarray}
y = 5\bm{\psi}_0(\bm{\xi}) + \bm{\psi}_1(\bm{\xi}) + \bm{\psi}_3(\bm{\xi}) + \bm{\psi}_4(\bm{\xi}) \nonumber
\end{eqnarray}

Transforming this expansion into a simple power series we obtain the folowing polynomial:
\begin{eqnarray}
y = 5 + \xi_1 + \xi_2 + \xi_1 \xi_2 \nonumber
\end{eqnarray}
where $a_{00} = 5, a_{10} = 1, a_{01} = 1$, and $a_{11} = 1$. Here the multi-index of maximum degree is $\mathbf{N} = (1,1)$.

The following intermediate coefficients are obtain from Eq.\eqref{interCoeff} in order to perform the scaling operation: $\tilde{a}_{00} = 4, \tilde{a}_{01} = 0, \tilde{a}_{10} = 0$, and $\tilde{a}_{11} = 1$. The final set of power-coefficients is given by Eq.\eqref{finalCoeff}: $\hat{a}_{00} = 4, \hat{a}_{01} = 0, \hat{a}_{10} = 0$, and $\hat{a}_{11} = 4$.

Finally, the Bernstein coefficients are obtain using Eq.\eqref{bernCoeff}: $\beta_{00} = 4, \beta_{01} = 4, \beta_{10} = 4$, and $\beta_{11} = 8$, and the Bernstein basis is given by:
\begin{eqnarray}
\mathbf{B}_{00}^{11} &= \frac{1}{16}(1-\xi_1)(1-\xi_2) \quad \mathbf{B}_{01}^{11} &= \frac{1}{16}(1-\xi_1)(\xi_2+1) \nonumber \\
\mathbf{B}_{10}^{11} &= \frac{1}{16}(\xi_1+1)(1-\xi_2) \quad \mathbf{B}_{11}^{11} &= \frac{1}{16}(\xi_1+1)(\xi_2+1) \nonumber
\end{eqnarray}
\subsection{Bounding the range of polynomials}
\label{sec:Range}
Given the Bernstein expansion in Eq.\eqref{bernExpansion}, the range enclosing property \cite{Garloff1993} gives a bound on the polynomial in terms of the Bernstein coefficients:
\begin{eqnarray}
\min_{\mathbf{I} \le \mathbf{N}} \beta_{\mathbf{I}} \le y(\bm\xi) \le \max_{\mathbf{J} \le \mathbf{N}} \beta_{\mathbf{J}} \quad \forall ~ \bm\xi \in G = [\underline{\bm\xi}, \overline{\bm\xi}]
\end{eqnarray}

Provided that the initial box is small enough, the range provided by the Bernstein form is exact. Compared with other forms in estimating the range, it is experimentally shown in Ref. \cite{Stahl1995} that the Bernstein form provides the smallest average overestimation error in the univariate case. For the previous example the range of $y$ is bounded by $[4,8]$. 

Tighter bounds can be obtained by subdivision of the initial box and choosing the minimum and the maximum of all the Bernstein coefficients corresponding to each sub-box. An efficient algorithm for range computation that incorporates a number of features such as subdivision, cut-off test, simplified vertex test, monotonicity test and others is provided in Ref.\cite{Ray2008}.

Therefore, getting back to our problem in mapping DS structures on closed intervals, Eq.\eqref{mapping} and Eq.\eqref{IntProp}, the output interval or the focal element $Y=[\underline{y},\overline{y}]$ is given by:
\begin{eqnarray}
\underline{y} = \min_{\mathbf{I} \le \mathbf{N}} \beta_{\mathbf{I}} \quad \mathrm{and} \quad \overline{y} = \max_{\mathbf{J} \le \mathbf{N}} \beta_{\mathbf{J}}
\end{eqnarray}

This methodology is applied to map all the focal elements in the initial body of evidence through the nonlinear function. Their corresponding masses are obtained using Eq.\eqref{mapping}. This way a body of evidence for the response is constructed.
\section{NUMERICAL RESULTS}
\label{sec:Results}
To prove the concept, we have selected an algebraic problem from a set of challenges used to investigate the propagation of epistemic uncertainty \cite{Oberkampf2004}. The presented problem has been investigated previously in the literature by Oberkampf and Helton \cite{Oberkampf2004a}. In the present paper we are using the exact parameters for the simulation as in Ref.\cite{Oberkampf2004a}.

Consider the following mapping:
\begin{eqnarray}
y = f(a,b) = (a+b)^a \label{myfun}
\end{eqnarray}
where the information concerning $a$ and $b$ is provided by the following sources and their corresponding bpa:
\begin{eqnarray}
\mathcal{A}_1 &:& \bigg\{ \big([0.6,0.9],1.0\big) \bigg\} \nonumber \\
\mathcal{A}_2 &:& \bigg\{ \big([0.1,0.5],0.2\big) ~,~ \big([0.5,1.0],0.8\big) \bigg\} \nonumber 
\end{eqnarray}
\small
\begin{eqnarray}
\mathcal{B}_1 &:& \bigg\{ \big([0.3,0.5],0.1\big) ~,~ \big([0.6,0.8],0.9\big) \bigg\} \nonumber \\
\mathcal{B}_2 &:& \bigg\{ \big([0.2,0.4],0.1\big) ~,~ \big([0.4,0.6],0.7\big) ~,~ \big([0.6,1.0],0.2\big) \bigg\} \nonumber \\
\mathcal{B}_3 &:& \bigg\{ \big([0.0,0.2],\frac{1}{3}\big) ~,~ \big([0.2,0.4],\frac{1}{3}\big) ~,~ \big([0.3,0.5],\frac{1}{3}\big) \bigg\} \nonumber
\end{eqnarray}
\normalsize

\begin{figure}[ht]
	\centering
		\includegraphics[width=3.5in]{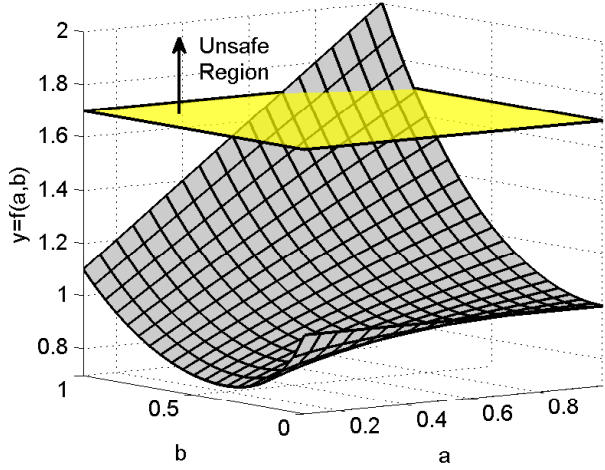}
	\caption{The function and the unsafe region}
	\label{fig:function}
\end{figure}

Given the above information, we are looking to bound the probability of the response in the unsafe region when $y > 1.7$. The function and the desired unsafe region are shown in Fig.\ref{fig:function}.

The bpa for $a$ and $b$ is obtained by aggregating the information from the first two sources and the last three sources respectively using the mixing rule in Eq.\eqref{mixingRule} under the equal reliability assumption. Thus the new DS structures obtained are given by:
\small
\begin{eqnarray}
\mathcal{A} &:& \bigg\{ \big([0.1,0.5],0.1\big) ~,~ \big([0.5,1.0],0.4\big) ~,~ \big([0.6,0.9],0.5\big) \bigg\} \nonumber \\
\mathcal{B} &:& \bigg\{ \big([0.0,0.2],0.111\big) ~,~ \big([0.2,0.4],0.144\big) ~,~ \nonumber \\ 
&&\quad \big([0.3,0.5],0.144\big) ~,~ \big([0.4,0.6],0.233\big)  ~,~ \nonumber \\ 
&&\quad \big([0.6,0.8],0.3\big) ~,~ \big([0.6,1.0],0.067\big) \bigg\} \nonumber
\end{eqnarray}
\normalsize

The focal elements of the DS structure $\mathcal{Y}$ are obtained by propagating the product space $\mathcal{A} \times \mathcal{B}$ through the nonlinear function in Eq.\eqref{myfun}, and the basic probability assignment is obtained using Eq.\eqref{mapping}. Thus the induced body of evidence for the response $y$ is given in Table \ref{tab:addlabel}. 

The following parameters have been used in obtaining the polynomial chaos expansion of the response and the afferent bounds: $n=2, p = 5$, and  $P = 19$ such that the total degree of the polynomial is no greater than $5$. The integrals in Eq.\eqref{Ycoeff} have been numerically evaluated using Gauss-Legendre quadrature rule with $20$ points in each direction, and the Bernstein coefficients have been obtain using $11$ subdivisions in each direction. The obtained intervals are compared with the intervals given by interval analysis (INTLAB \cite{Rump1999}), and with the reference intervals provided by a genetic algorithm (GA).

In Table \ref{tab:addlabel}, given the $i$th focal element of $\mathcal{A}$ and the $j$th focal element of $\mathcal{B}$, the Box\# is given by $3(j-1)+i$. The numbers in bold indicate that a smaller lower bound or a larger upper bound has been found by the genetic algorithm. In this particular example we have overestimated most of the lower bounds and we have provided no underestimation for the upper bounds.

\begin{figure}[ht]
	\centering
		\includegraphics[width=3.5in]{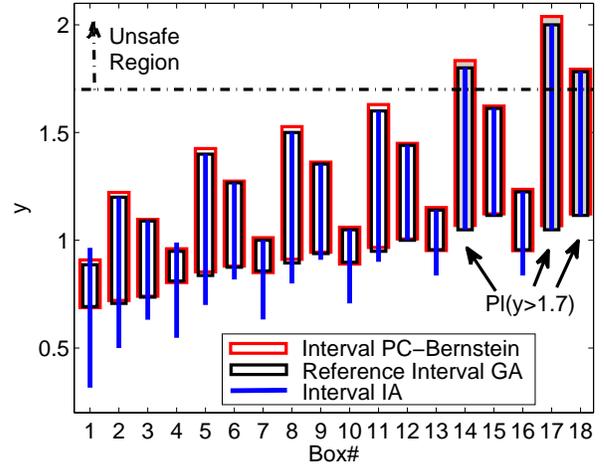}
	\caption{Focal elements from the induced DS structure}
	\label{fig:boxes}
\end{figure}

The focal elements from the DS structure in Table \ref{tab:addlabel} are also graphically presented in Fig.\ref{fig:boxes}. The wider boxes represent the bounds found by the proposed approach while the narrow boxes depict the bounds returned by the genetic algorithm, and the lines show the range computed using interval arithmetics.
%
\begin{table}[htbp]
  \centering
  \caption{Induced DS structure for the response $y$}
    \begin{tabular}{rrrrrrrrr}
    \addlinespace
    \toprule
     Box\#    &   $\underline{y}$~~~     &    $\overline{y}$~~~   &   $m_f~$    &       &   Box\#   &    $\underline{y}$~~~    &   $\overline{y}$~~~    &  $m_f~$ \\
    \midrule
    1     & 0.687 & 0.909 & 0.011 &       & 10    & 0.890 & 1.061 & 0.023 \\
    2     & \textbf{0.721} & 1.222 & 0.044 &       & 11    & \textbf{0.967} & 1.630 & 0.093 \\
    3     & \textbf{0.741} & 1.097 & 0.056 &       & 12    & \textbf{1.007} & 1.450 & 0.117 \\
    4     & 0.804 & 0.961 & 0.014 &       & 13    & 0.953 & 1.152 & 0.030 \\
    5     & \textbf{0.853} & 1.426 & 0.058 &       & 14    & \textbf{1.069} & 1.834 & 0.120 \\
    6     & \textbf{0.880} & 1.275 & 0.072 &       & 15    & \textbf{1.123} & 1.623 & 0.150 \\
    7     & 0.850 & 1.012 & 0.014 &       & 16    & 0.952 & 1.236 & 0.007 \\
    8     & \textbf{0.912} & 1.528 & 0.058 &       & 17    & \textbf{1.068} & 2.039 & 0.027 \\
    9     & \textbf{0.945} & 1.363 & 0.072 &       & 18    & \textbf{1.122} & 1.794 & 0.033 \\
    \bottomrule
    \end{tabular}
  \label{tab:addlabel}
\end{table}

To compute the lowest and the highest probability of $y>1.7$, we use the Fig.\ref{fig:boxes} to sum over all the bpa's of the intervals that are properly included in the unsafe region for the lower bound and for the upper bound, sum over all the bpa's of the intervals that intersect the unsafe region. No intervals are properly included in the unsafe region, thus the lower bound is $0.0$. Three intervals are found to intersect the unsafe region, boxes: $14$, $17$ and $18$. Summing over their bpa's we found the upper bound to be $0.18$. All three methods give $0.0 \le Prob(y > 1.7) \le 0.18$, which is in agreement with the result published by Oberkampf in Ref.\cite{Oberkampf2004a}.

\begin{figure}[ht]
	\centering
		\includegraphics[width=3.5in]{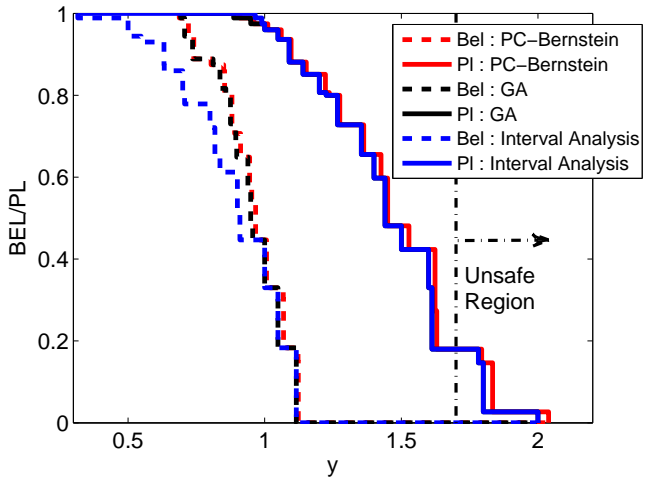}
	\caption{CCBF and CCPF from the induced DS structure}
	\label{fig:ccdf}
\end{figure}

Both the CCBF and the CCPF given by Eq.\eqref{ccdf1}-\eqref{ccdf2} are plotted in Fig.\ref{fig:ccdf} along with the marking for the unsafe region. A reason of concern is the overestimation of the lower bound, due to the finite polynomial chaos expansion, which in this example may provide a larger lower bound for the  probability of failure. Observe the gross bounds provided by the interval arithmetics due to dependency effect.
\section{CONCLUSIONS}
\label{sec:Conclusions}
A new approach in approximating the interval propagation for epistemic uncertainty quantification has been presented. The input variables are represented as a polynomial expansion of random variables on compact support, and by applying the Galerkin projection in a non-intrusive way we find the response of the system also as a polynomial expansion, here in the Legendre basis. It exploits the efficient mapping of random variables using polynomial chaos expansion from which it extracts only the bounds of its compact support.

We further propose to transform the output polynomial chaos expansion from the Legendre basis to the Bernstein basis, and use the range enclosure property of Bernstein forms to efficiently extract the bounds of the range of the output. The method does not suffer of the dependency problem as in the interval arithmetics, however much work remains to be done in studying the accuracy of interval propagation using polynomial chaos. The proposed approach is applied on a challenge problem previously investigated by Oberkampf et al. to propagate epistemic uncertainty and the numerical results obtained provide a basis for optimism.
\bibliographystyle{plain}
\bibliography{IPpaper}

\end{document}